\documentclass[a4paper,11pt]{article}
\usepackage{pos}
\usepackage{graphicx}
\usepackage{subcaption}
\usepackage{lipsum}
\usepackage{amsmath}
\usepackage[normalem]{ulem}
\usepackage{natbib}
\usepackage{bigints}
\let\OLDthebibliography\thebibliography
\renewcommand\thebibliography[1]{
  \OLDthebibliography{#1}
  \setlength{\parskip}{0pt}
  \setlength{\itemsep}{-1pt}
}

\title{Asymmetric collisions in \texttt{MadGraph5\_aMC@NLO}}

\author*[a]{Laboni Manna}
\author[a]{Anton Safronov}
\author[b,c]{Carlo Flore}
\author[a]{Daniel Kikola}
\author[d]{Jean-Philippe Lansberg}
\author[e]{Olivier Mattelaer}

\affiliation[a]{Warsaw University of Technology,\\
plac Politechniki 1, Warsaw, Poland}
\affiliation[b]{INFN, Sezione di Torino,\\
Via P. Giuria 1, Torino I-10125, Italy}
\affiliation[c]{Dipartimento di Fisica, Università di Torino,\\
Via P. Giuria 1, Torino I-10125, Italy}
\affiliation[d]{Université Paris-Saclay,\\
CNRS, ĲCLab, 91405 Orsay, France}
\affiliation[e]{Centre for Cosmology, Particle Physics and Phenomenology (CP3),\\
Université Catholique de Louvain, Chemin du Cyclotron, Louvain-la-Neuve, B-1348, Belgium,}

\emailAdd{laboni.manna.dokt@pw.edu.pl}
\emailAdd{anton.safronov.dokt@pw.edu.pl}
\emailAdd{carlo.flore@unito.it}
\emailAdd{daniel.kikola@pw.edu.pl}
\emailAdd{Jean-Philippe.Lansberg@in2p3.fr}
\emailAdd{ olivier.mattelaer@uclouvain.be}

\abstract{We will gain unprecedented, high-accuracy insights into the internal structure of the atomic nucleus thanks to lepton-hadron collision studies in the coming years at the Electron-Ion-Collider (EIC) in the United States. A good control of radiative corrections is necessary for the EIC to be fully exploited and to extract valuable information from various measurements. We present our extension of photoproduction at fixed order in \texttt{MadGraph5\_aMC@NLO}, a widely used framework for (next-to-)leading order calculations at the Large Hadron Collider (LHC). It applies to electron-hadron collisions, in which the quasi-real photon comes from an electron as well as to proton-nucleus and nucleus-nucleus collisions.}

\FullConference{The European Physical Society Conference on High Energy Physics (EPS-HEP2023)\\
 21-25 August 2023\\
Hamburg, Germany\\}

\begin{document}
\maketitle

\vspace*{-0.3cm}
\section{Introduction}
To delve deeper into the proton internal composition and that of the nucleus, Brookhaven National Laboratory in the US is going to build the Electron-Ion Collider (EIC)~\cite{AbdulKhalek:2021gbh}. This cutting-edge facility promises to unveil previously elusive facets of proton and nuclear structure, opening a new portal into the enigmatic realm of the atomic nucleus.

In order to effectively accomplish future measurements, optimize detector performances, and execute data collection campaigns at the EIC, a trustworthy simulation tool for electron-proton ($ep$) collisions is mandatory. While several tools, such as \texttt{HELAC-Onia}~\cite{Shao:2015vga}, \texttt{Pythia}~\cite{Helenius:2017aqz} or single-usage codes such as \texttt{FMNR}~\cite{Frixione:1993dg}, are currently available, they also have limitations. Some are restricted up to leading-order (LO) in $\alpha_{s}$, while others, like \texttt{FMNR}, are not automated. The development of reliable event generators for photoproduction is therefore important, as recently done for \texttt{SHERPA}~\cite{Hoeche:2023gme,Meinzinger:2023xuf}, for the upcoming EIC. 

In this work, we will illustrate our extension and validation of photoproduction at next-to-leading order (NLO) in \texttt{MadGraph5\_aMC@NLO} (\texttt{MG5})~\cite{Alwall:2014hca}. \texttt{MG5} has the capability to automatically compute NLO results. As NLO computations provide a more comprehensive depiction of spectra with smaller uncertainty compared to LO ones, they represent an invaluable asset for our research at the EIC. In what follows, we will show the capabilities of our new tool for the production of heavy quarks. Among these studies, inclusive photoproduction of open charm and bottom quarks has significant importance, which will advance our understanding of perturbative quantum chromodynamics (pQCD). In addition, we will highlight future possibilities. 

\vspace*{-0.3cm}
\section{Framework}
According to the collinear QCD factorization theorem, the cross section for the scattering of two hadrons producing anything ($X$) in the final state can be written as a convolution of a perturbatively calculable partonic cross section and non-perturbative parton distribution functions (PDFs) of hadrons:
\begin{equation}
    \sigma_{AA\rightarrow X}=\sum\limits_{i,j} \int\!dx_{i} dx_{j}{f_{i}^{A} (x_{i},\mu_{F}; {\tt LHAID}) }{f_{j}^{A} (x_{j},\mu_{F}; {\tt LHAID}) }{\hat{{\sigma}}}_{ab \rightarrow X}(x_{i},x_{j},\mu_{F},\mu_{R})
    \label{eq:equation1}
\end{equation}
where $x_{i,j}$ are the momentum fractions carried by the partons (gluon or quark) from the hadrons, $\mu_{R, F}$ are the renormalisation and factorisation scale respectively, $f_{i}^{A}$, $f_{j}^{A}$ are the PDFs of incoming hadrons and $\hat{{\sigma}}_{ab \rightarrow X}$ is the partonic cross section for the process.  Eq.~(\ref{eq:equation1}) is the fundamental equation upon which \texttt{MG5} has been developed and is specific for symmetric \textit{AA} collisions with the same \texttt{LHAID}. As an extension of \texttt{MG5}, we have included two different types of asymmetric collisions: asymmetric hadron-hadron collisions and electron-hadron collisions (photoproduction). In the first scenario, after having modified the existing algorithm of \texttt{MG5}, we can call simultaneously two distinct LHAPDF sets and calculate the corresponding cross section as:

\begin{equation}
    \sigma_{AB\rightarrow X}=\sum\limits_{i,j} \int\!dx_{i} dx_{j}{f_{i}^{A} (x_{i},\mu_{F}; {\tt LHAID1}) }{f_{j}^{B} (x_{j},\mu_{F}; {\tt LHAID2}) }{\hat{{\sigma}}}_{ab \rightarrow X}(x_{i},x_{j},\mu_{F},\mu_{R})
    \label{eq:equation2}
\end{equation}

More details about the implementation can be found in Ref.~\cite{Safronov:2022uuy}. In the photoproduction\footnote{Photoproduction occurs in $ep$ collision when the electron scattered at a small angle and Q$^{2}_{\rm max} \leq\mathcal{O}(1 $ GeV$^{2})$.} case, one needs to replace one of the PDFs in Eq.~(\ref{eq:equation1}) with the photon flux $f_{\gamma}^{e} (x_{\gamma}, Q^{2}_{\rm max})$\footnote{Here we have used equivalent photon approximation (EPA)~\cite{Frixione:1993yw}.} and compute the cross section. We have two relevant contributions for photoproduction: direct and resolved. Our work focuses on the extension to direct photoproduction\footnote{Though it depends on the value of $\sqrt{s}$ and inelasticity $z$, heavy-flavour photoproduction is dominated by the direct photoproduction.}. The total inclusive electron-hadron cross section for direct photoproduction can be written as~\cite{Toll:2011tm}

\begin{equation}
    \sigma_{eh\rightarrow X}=\sum\limits_{j} \int\!dx_{\gamma} dx_{j}{f_{\gamma}^{e} (x_{\gamma},Q^{2}_{\rm max}) }{f_{j}^{h} (x_{j},\mu_{F};{\tt LHAID}) }{\hat{{\sigma}}}_{\gamma j \rightarrow X}(x_{\gamma},x_{j},\mu_{F},\mu_{R})
\end{equation}

where $x_{\gamma}$ is the momentum fraction carried by the photon from the electron and $Q^2_{\rm max}$ is the maximal photon virtuality. To enable photoproduction within \texttt{MG5}, we have introduced two distinct variables, one governing $Q^2_{\rm max}$ and the other controlling the factorisation scale. This extension required meticulous adjustments to \texttt{MG5} routines for efficient operations and now provides correct control over $Q^2_{\rm max}$, which is an experimental parameter as opposed to $\mu_{F}$. 

\vspace*{-0.3cm}
\section{Validation}
In order to validate our \texttt{MG5} extension to direct photon production, we have compared our results with theoretical predictions for $b$ and $c$ quarks photoproduction in $ep$ collisions at HERA, both at LO and NLO in pQCD.

\vspace*{-0.3cm}
\begin{figure}[htbp]
\centering
  \includegraphics[width=.49\textwidth]{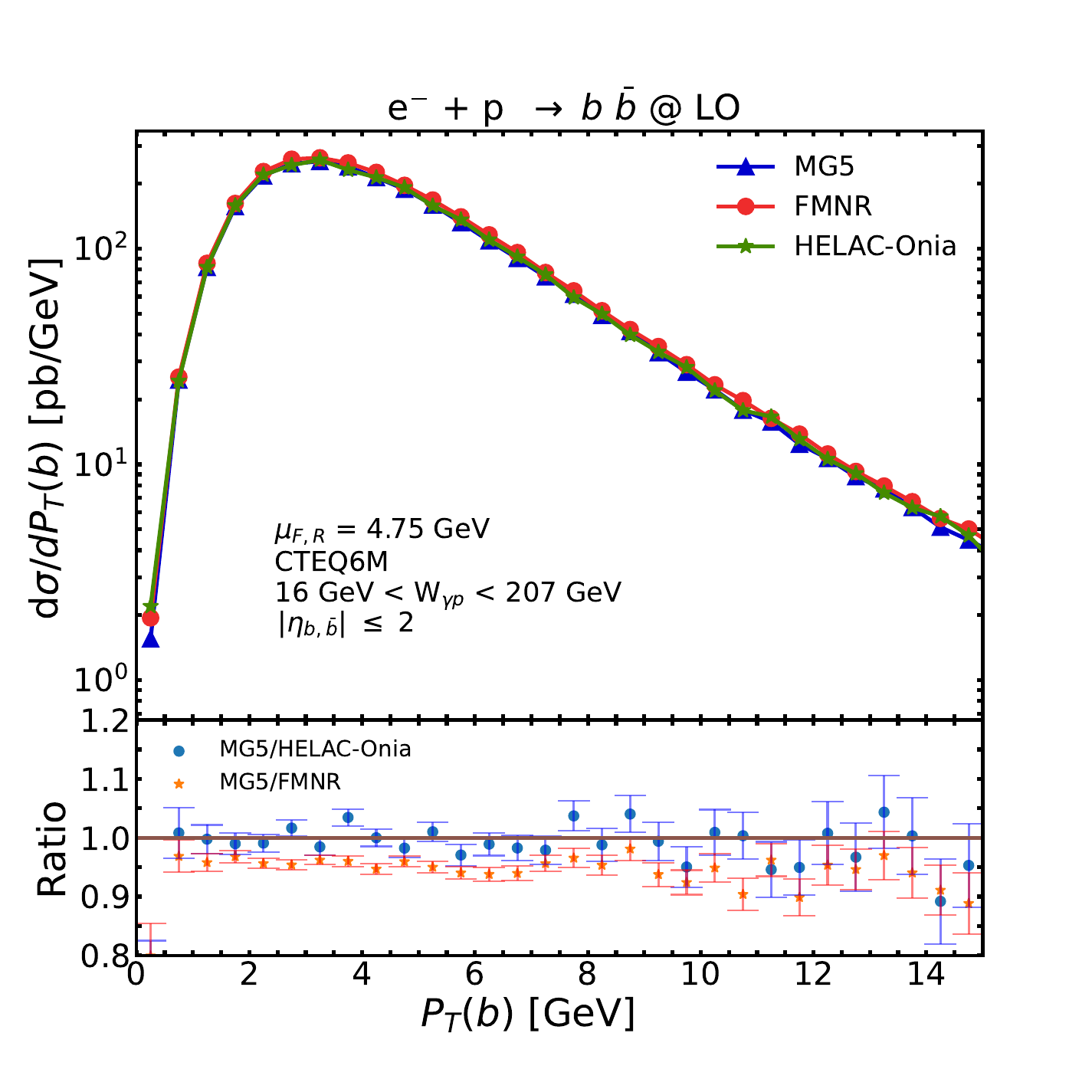}
  \includegraphics[width=.49\textwidth]{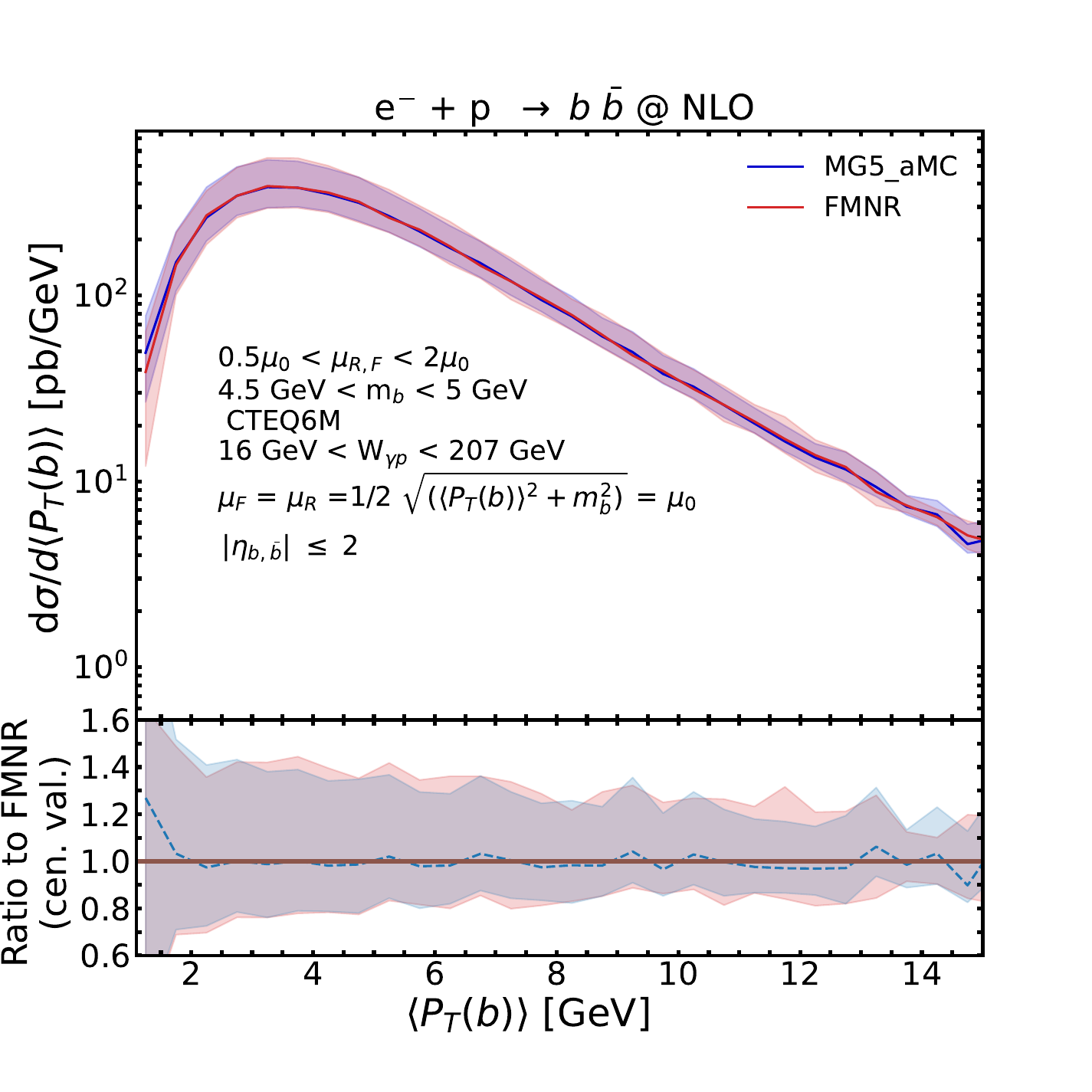}
\vspace*{-0.3cm}
\caption{Comparison of the transverse momentum distribution of bottom quark photoproduction by \texttt{MG5} at $\sqrt{s}$ = 319 GeV with results at LO from \texttt{HELAC-Onia} and \texttt{FMNR} (left) and at NLO from \texttt{FMNR} (right).}
\label{fig1}
\end{figure}
In Fig.~\ref{fig1} (left panel), we present the transverse momentum ($P_{T} (b)$) distribution of $b$ quark photoproduction at LO. Specifically, we compare the results obtained from our \texttt{MG5} extension with those generated by \texttt{HELAC-Onia} and the \texttt{FMNR} program. For consistency across all the event generators\footnote{Since FMNR does not allow one to use LHAPDF6.}, we have adopted the CTEQ6M PDF set~\cite{Nadolsky:2008zw}, with $\mu_F = \mu_R = m_b$ = 4.75 GeV being the mass of the $b$ quark. The plot clearly illustrates an agreement up to about 5\% for the P$_{T}(b)$ distributions at LO, underscoring the results from the different event generators (\texttt{HELAC-Onia} and \texttt{FMNR}).

{Our study also validates NLO computations against results published in Ref.~\cite{H1:2012ffl},  in which both resolved and direct photoproduction were considered, while we focus solely on the direct process. In Fig.~\ref{fig1} (right panel)}, we present the quadratically averaged transverse momentum $\langle P_{T}(b)\rangle$ distribution, where $\langle P_{T}(b)\rangle$ = $\sqrt{(P_{T}(b)^{2} + P_{T}(\Bar{b})^{2})/2 }$ for $b$-quark production at a center-of-mass (CM) energy of $\sqrt{s} = 319$ GeV.

{In this computation, we have systematically explored the sensitivity to $m_b$, varying it in the range $[4.5 \div 5.0]$ GeV. We have set $\mu_{R,F} = \mu_{0} = \tfrac12 \sqrt{m_{b}^{2}+\langle P_{T}(b)\rangle^{2}}$, and the corresponding uncertainty was evaluated at the extreme points of $0.5\mu_{0} < \mu_{R,F} < 2\mu_{0}$. We applied the following kinematical cuts: $\lvert \eta_{b, \bar{b}}\rvert \leq 2$ and $16$ GeV $ < W_{\gamma p } < 207$ GeV, with $W_{\gamma p}$ }being the photon-proton CM energy. The combined uncertainty, as shown in Fig.~\ref{fig1} (right panel), is the sum in quadrature of maximum and minimum values arising from mass and scale variations. Our results demonstrate an agreement up to $\sim\mathcal{O}$(1\%) between \texttt{FMNR} and \texttt{MG5} for $P_T(b) > 1$ GeV for direct photoproduction. We also have a comparable agreement for the charm production.

\vspace*{-0.3cm}
\section{Predictions}
With the successful validation of photoproduction in \texttt{MG5}, we are now well-equipped to extend our investigations to various planned $ep$ facilities characterized by different CM energies, such as the Electron-Ion Collider China (EIcC)~\cite{Anderle:2021wcy}, the Large Hadron Electron Collider (LHeC)~\cite{LHeCStudyGroup:2012zhm}, and the Future Circular lepton-hadron Collider (FCC-eh)~\cite{Narain:2022qud}. These facilities will offer an impressive spectrum of CM energies, catering to a variety of research objectives.
\begin{table}[htbp]\footnotesize
    \centering
    \begin{tabular}{|c|c|c|c|c|}
    \hline
       $ep$ facilities & EIcC & EIC & LHeC & FCC-eh \\
        \hline
         $\sqrt{s}$& 16.7 GeV  & 45-140 GeV  & 1.2 TeV  & 3.4 TeV \\
         \hline 
         Luminosity & 50 fb$^{-1}$  & 10-100  fb$^{-1}$  & 100  fb$^{-1}$ & 100  fb$^{-1}$\\
         \hline
    \end{tabular}
    \caption{CM energies for different future electron-proton experiments.}
    \label{tab:my_label}
\end{table}
\begin{figure}[htbp]
 \centering
  \includegraphics[width=.49\textwidth]{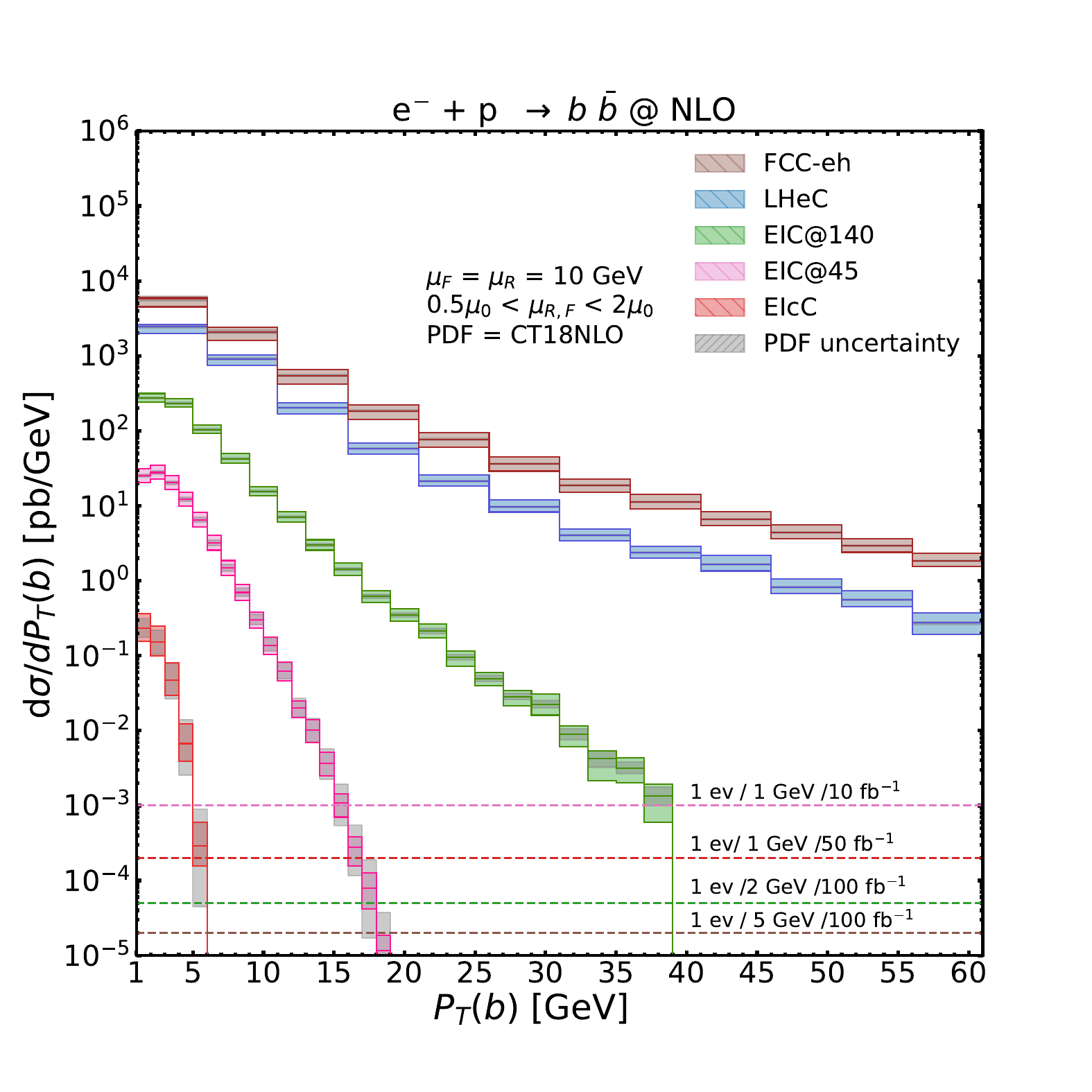}\hspace*{-.5cm} 
  \includegraphics[width=.49\textwidth]{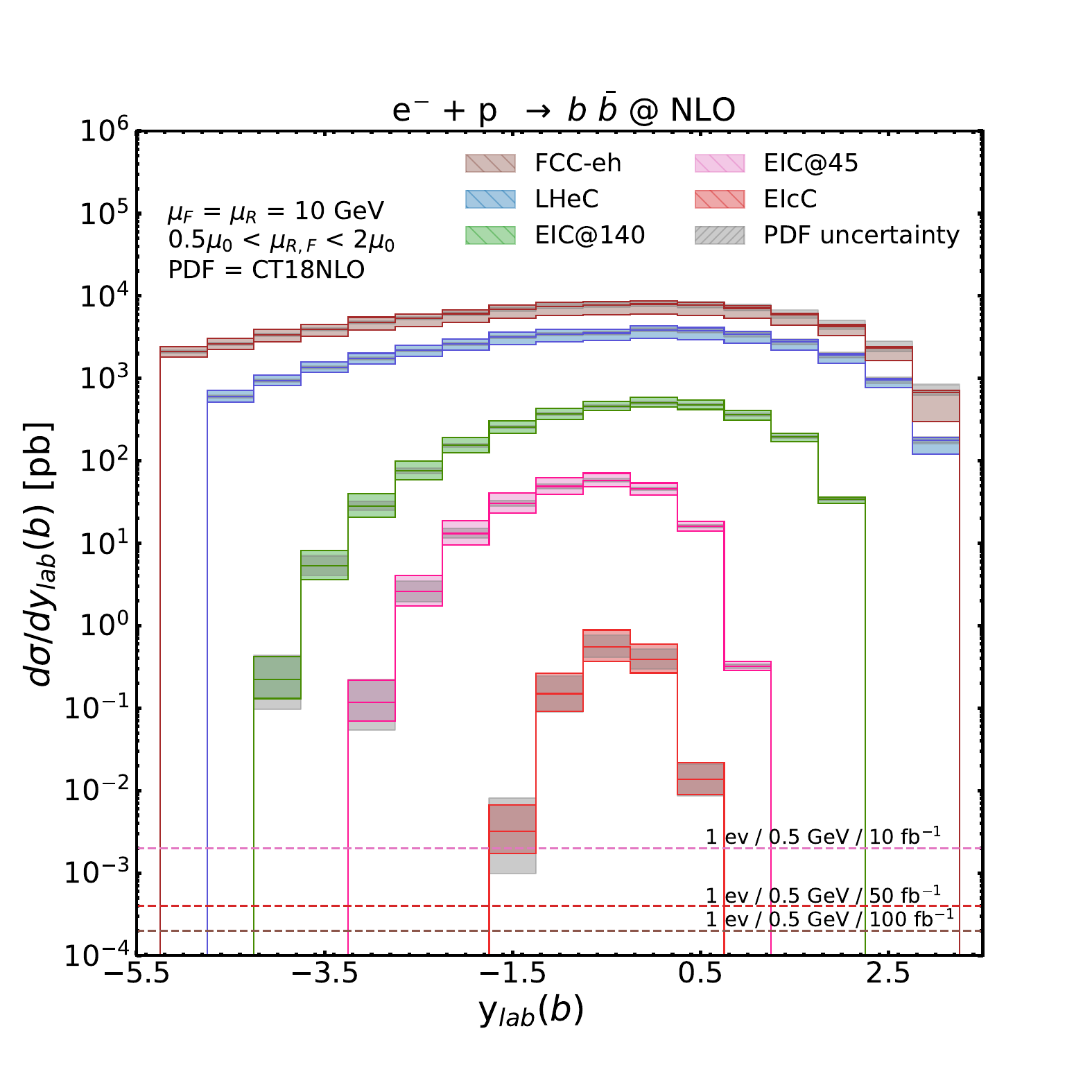}
\vspace*{-0.3cm}
\caption{Predictions by \texttt{MG5} for transverse momentum (left) and rapidity distribution (right) for $b$-quark photoproduction in $ep$ collisions at different values of the CM energy.}
\label{fig2}
\end{figure}
Leveraging our extension in \texttt{MG5}, we have conducted predictive studies encompassing the NLO transverse momentum and rapidity distributions for both $b$ and $c$ quark production for the aforementioned experiments. Here, we focus on the $b$ quark photoproduction. Results are presented in Fig.~\ref{fig2}. We have fixed $\mu_F = \mu_R = 10$ GeV, with the corresponding scale uncertainties spanning the range 0.5 $\mu_{0}$ <  $\mu_{R,F}$ < 2 $\mu_{0}$. Adopting the CT18NLO PDF set~\cite{Hou:2019efy}, we show the PDF uncertainties via grey bands, and scale uncertainty bands for each CM energy. Horizontal observability lines, accompanied by their respective integrated luminosities and with a 10\% detection efficiency for the bottom quark, are also shown, allowing us to predict \textit{P$_{T}$} or rapidity ranges within which the $b$ quark can be reliably detected. {Similar results can also be achieved for charm photoproduction as well}.

\vspace*{-0.3cm}
\section{Conclusions}
In summary, we have discussed the inclusion of direct photoproduction processes within \texttt{MG5}. This new extension empowers us to predict a wide range of observables, including rapidity and transverse momentum, for both charm- and bottom-quark production at any CM energy. While our current extension focuses on direct photoproduction, it offers the potential for expansion to include resolved processes, as it can use two distinct PDFs~\cite{Safronov:2022uuy}. Furthermore, our versatile tool extends its applicability to explore exotic processes such as top quark and Higgs boson photoproduction. Beyond photoproduction processes in electron-proton collisions, our extension opens the door for theoretical studies of \textit{inclusive} ultra-peripheral collisions at the LHC~\cite{Baltz:2007kq} and other hadronic colliders. Our current extension is not publicly accessible at the moment, but we are actively working towards its forthcoming release that will soon be accessible through the EU Virtual Access platform NLOAccess~\cite{Flore:2023dps} (\url{https://nloaccess.in2p3.fr}). We anticipate that this enhanced extension will play a crucial role in advancing both theoretical predictions and the analysis of experimental data for forthcoming experiments at $ep$ and $eA$ facilities.

\vspace*{-0.3cm}
\section*{Acknowledgements}\vspace*{-0.3cm}
This work was supported in part by the Excellence Initiative: Research University at Warsaw University of Technology and the European Union’s Horizon 2020 research and innovation program under Grant Agreements No. 824093 (Strong2020) in order to contribute to the EU Virtual Access ``NLOAccess'' and no. 722104 as part of the Marie Sk\l odowska-Curie Innovative Training Network MCnetITN3. This project has also received funding from the French Agence Nationale de la Recherche (ANR) via the grant ANR-20-CE31-0015 (``PrecisOnium'')  and was also partly supported by the French CNRS via the COPIN-IN2P3 bilateral agreement. 

\bibliographystyle{utphys}
\bibliography{reference}

\end{document}